\documentclass[useAMS, usenatbib, usegraphicx]{mn2e}
\usepackage{times}
\usepackage{color}
\usepackage{amssymb}
\usepackage{amsmath}
\usepackage{enumerate}
\usepackage{upgreek}

\title[SFE in spiral arms]{Star formation scales and efficiency in Galactic spiral arms}
\author[D. J. Eden et al.]{D. J. Eden$^{1,2}$\thanks{E-mail: david.eden@astro.unistra.fr}, T. J. T. Moore$^{2}$, J.S. Urquhart$^{3}$, D. Elia$^{4}$, R. Plume$^{5}$, A.J. Rigby$^{2}$ \newauthor  \& M.A. Thompson$^{6}$\\
$^{1}$Observatoire astronomique de Strasbourg, Universit\'{e} de Strasbourg, CNRS, UMR 7550, 11 rue de l'Universit\'{e}, 67000, Strasbourg, France\\
$^{2}$Astrophysics Research Institute, Liverpool John Moores University, IC2, Liverpool Science Park, 146 Brownlow Hill, Liverpool L3 5RF\\
$^{3}$Max-Planck-Institut f\"{u}r Radioastronomie, Auf dem H\"{u}gel 69, 53121 Bonn, Germany\\
$^{4}$Istituto di Astrofisica e Planetologia Spaziali - INAF, Via Fosso del Cavaliere 100, I-00133 Roma, Italy\\
$^{5}$Department of Physics and Astronomy, University of Calgary, 2500 University Drive NW, Calgary, Alberta T2N 1N4, Canada\\
$^{6}$Centre for Astrophysics Research, Science \& Technology Research Institute, University of Hertfordshire, College Lane, Hatfield, Herts AL10 9AB}
\begin{document}

\date{Accepted. Received; in original form}

\pagerange{\pageref{firstpage}--\pageref{lastpage}} \pubyear{2015}

\maketitle

\label{firstpage}

\begin{abstract}

We positionally match a sample of infrared-selected young stellar objects (YSOs), identified by combining the $\emph{Spitzer}$ GLIMPSE, WISE and $\emph{Herschel Space Observatory}$ Hi-GAL surveys, to the dense clumps identified in the millimetre continuum by the Bolocam Galactic Plane Survey in two Galactic lines of sight centred towards $\emph{l}$ = 30$\degr$ and $\emph{l}$ = 40$\degr$. We calculate the ratio of infrared luminosity, $L_{\rmn{IR}}$, to the mass of the clump, $M_{\rmn{clump}}$, in a variety of Galactic environments and find it to be somewhat enhanced in spiral arms compared to the interarm regions when averaged over kiloparsec scales. We find no compelling evidence that these changes are due to the mechanical influence of the spiral arm on the star-formation efficiency rather than, e.g., different gradients in the star-formation rate due to patchy or intermittent star formation, or local variations that are not averaged out due to small source samples. The largest variation in $L_{\rmn{IR}}/M_{\rmn{clump}}$ is found in individual clump values, which follow a log-normal distribution and have a range of over three orders of magnitude. This spread is intrinsic as no dependence of $L_{\rmn{IR}}/M_{\rmn{clump}}$ with $M_{\rmn{clump}}$ was found. No difference was found in the luminosity distribution of sources in the arm and interarm samples and a strong linear correlation was found between $L_{\rmn{IR}}$ and $M_{\rmn{clump}}$.

\end{abstract}


\section{Introduction}

It is clear from many images of ``grand design'' spiral galaxies that star formation and spiral arm topology are closely interlinked. However, the precise relationship between the physics of the spiral arms and the nature of their ensuing star formation is not yet clear. The influence on the spiral arms could either be that they are triggering one, or more, of the main evolutionary stages of star formation \citep{Roberts69}, or are organising features for the star-forming material (\citealt{Elmegreen86}; \citealt*{Vogel88}). The main stages of star formation are represented by: the conversion of atomic gas to molecular clouds, molecular clouds with internal dense clumps and the clumps forming cores from which the single stars form.

The evidence of the first of these potential influences is ambiguous. Increases in the efficiency of cloud formation from the neutral interstellar medium have been both observed \citep{Heyer98} and predicted \citep*{Dobbs06} but the results of \citet{Foyle10} contradict this. They find no increase in that efficiency, or the star-formation efficiency (SFE) in or out of the spiral arms of two spiral galaxies, NGC 628 and NGC 5194. There is also no observed difference in the fraction of the total molecular gas mass in the form of dense, star-forming clumps within clouds located in and out of spiral arms in the Milky Way, when averaged over kiloparsec scales \citep{Eden12, Eden13}. There are, however, large cloud-to-cloud variations in the dense-clump fraction. There are large increases in the star-formation-rate density ($\Sigma_{\rmn{SFR}}$ in units of M$_{\odot}$\,yr$^{-1}$\,kpc$^{-2}$) by factors of between $\sim$\,3 and $\sim$\,30 associated with spiral arms in the Galaxy. However, no more than 30 per cent of this results from increases in the integrated luminosity of the YSOs per unit molecular gas mass, while the rest is due to source crowding. Further, much of the residual increase in $L/M$ can be ascribed to the inclusion of extreme star-forming complexes such as W49 and W51 in the samples \citep{Moore12}.

This source crowding suggests that the dominant effect is due to the organising nature of the spiral arms. However, \citet*{Dobbs11} found that the spiral arms allow larger molecular clouds to form, by delaying clouds in their orbits, which may affect the mass function of clusters that form within them \citep*[e.g.][]{Weidner10}. The star-formation rate or efficiency in the higher-column density clouds are not increased, however \citep{Krumholz10}. The spiral arms may also be prolonging the lifetime of the clouds, lengthening the timescale for star formation to occur, increasing the star-formation efficiency (SFE). A study using the Galactic Ring Survey (GRS; \citealt{Jackson06}) found that the clouds in the interarm regions dissipate quickly \citep{Roman-Duval09} with models also predicting this \citep{Dobbs13}.

The aim of this study is to investigate the efficiency with which the dense clumps form stars and measure the variation of that efficiency from clump to clump and on kiloparsec scales, looking for changes associated with environment. To do this, it is necessary to use infrared wavelengths that trace the YSOs embedded within the clump. We therefore aim to combine infrared surveys that trace the embedded stellar population of YSOs and the Bolocam Galactic Plane Survey (BGPS; \citealt{Aguirre11}) to calculate the ratio of the infrared luminosity to the clump mass, an analogue of the SFE, as a function of Galactocentric radius and Galactic environment. This stage of the hierarchical, evolutionary star-formation process (which converts atomic gas to molecular clouds to dense clumps to YSOs) is the last to be investigated. Our previous work has found factors of $\sim$ 2-3 increases in the ratio of YSO luminosity to molecular-cloud mass, in clouds associated with spiral arms, but no enhancement in the average dense gas mass fraction (\citealt{Moore12}; \citealt{Eden12,Eden13}). We would therefore expect to find similar factors of $\sim$ 2-3 increases in the average clump to YSO conversion efficiency, measured by L$_{\rmn{IR}}$/M$_{\rmn{clump}}$, on similar scales, although the results may be sample-dependent in detail.

The two slices of the plane studied are the same as those outlined in \citet{Eden12, Eden13}, the $\emph{l}$\,=\,30$\degr$ region spanning $\emph{l}$\,=\,28$\overset{\circ}{.}$50\,--\,31$\overset{\circ}{.}$50 and the $\emph{l}$\,=\,40$\degr$ region covering Galactic longitudes of $\emph{l}$\,=\,37$\overset{\circ}{.}$83\,--\,42$\overset{\circ}{.}$50. The $\emph{l}$\,=\,30$\degr$ region extends to Galactic latitudes of $\mid$\emph{b}$\mid$\,$\leq$\,1$\degr$, whereas the $\emph{l}$\,=\,40$\degr$ region only covers latitudes of $\mid$\emph{b}$\mid$\,$\leq$\,0$\overset{\circ}{.}$5. The $\emph{l}$\,=\,30$\degr$ region has three populations within it: foreground material which is possibly associated with the Sagittarius arm, the tangent region which is the tangent of the Scutum--Centaurus arm and the end of the Long Bar, which hosts the extreme star-forming region W43 \citep{NguyenLuong11}, and the background spiral arms, which are the Sagittarius and Perseus arms. The $\emph{l}$\,=\,40$\degr$ region contains two passes of the Sagittarius arm, the outer edge of the Scutum--Centaurus tangent and the Perseus arm, as well as the interarm regions between the spiral arms.

A four-armed model of the spiral arm is assumed, as refined by multiple tracers, from H\,{\sc ii} regions (e.g. \citealt{Georgelin76}; \citealt*{Paladini04}) to the H\,{\sc i} distribution \citep*{Levine06} to combinations of tracers (e.g. \citealt*{Hou09}; \citealt{Urquhart13a, Urquhart13b}). The four spiral arms are the Norma, Scutum--Centaurus, Sagittarius and Perseus arms.

In Section 2, we outline the data sets which are used for this project, with Section 3 detailing the reasoning behind the wavelength choices and the calculation of YSO luminosities, whilst in Section 4 we present the results and analysis and Section 5 is a discussion of the results. A summary and the concluding remarks follow in Section 6.

\section{Data Sets}

\subsection{Infrared data sets}

The Galactic Legacy Infrared Mid-Plane Survey Extraordinaire (GLIMPSE) was a $\emph{Spitzer Space Telescope}$ Legacy Program, which (and with subsequent follow-ups) mapped two-thirds of the Galactic plane as part of three separate surveys \citep{Benjamin03, Churchwell09}. The $\emph{l}$\,=\,30$\degr$ and $\emph{l}$\,=\,40$\degr$ regions fall in the section observed during GLIMPSE I, which covered the longitudes of $\mid$\emph{l}$\mid$\,=\,10$\degr$\,--\,65$\degr$. The survey mapped this region at wavelengths of 3.6, 4.5, 5.8 and 8.0\,$\upmu$m using the Infrared Array Camera (IRAC; \citealt{Fazio04}) at spatial resolutions of $<$\,2$\arcsec$. The sensitivity limits at each wavelength were 15.5, 15.0, 13.0 and 13.0\,magnitudes, respectively with the high reliability ($\geq$ 99.5 per cent completeness) Point Source Catalogue containing $\sim$ 30 million sources \citep{Churchwell09}.

The Wide-field Infrared Survey Explorer (WISE; \citealt{Wright10}) mapped the whole sky at 3.4, 4.6, 12 and 22\,$\upmu$m with angular resolutions of 6.1$\arcsec$, 6.4$\arcsec$, 6.5$\arcsec$ and 12$\arcsec$, respectively. 5-$\sigma$ point-source sensitivities of 0.08, 0.11, 1 and 6\,mJy were achieved. A saturation limit of 330\,Jy is found for the brightest sources at 22\,$\upmu$m, but this is considerably deeper than MSX at 21\,$\upmu$m, $\sim$ 2.7\,Jy (\citealt{Price01}; \citealt{Lumsden13}).

Hi-GAL (Herschel infrared Galactic Plane Survey; \citealt{Molinari10, Molinari10a}), a $\emph{Herschel Space Observatory}$ Open-time Key Project, has mapped the entire Galactic plane at 70, 160, 250, 350 and 500\,$\upmu$m at spatial resolutions of 4$\arcsec$\,--\,40$\arcsec$ using the PACS \citep{Poglitsch08} and SPIRE \citep{Griffin08} cameras. The 90 per cent completeness limits at each wavelength are 0.5, 4.1, 4.1, 3.2 and 2.5\,Jy at the five wavelengths, respectively \citep{Molinari10}, for the $\emph{l}$\,=\,30$\degr$ region.

A custom source extraction algorithm, $\texttt{CuTEx}$ (Curvature Thresholding Extractor; \citealt{Molinari11}), was designed to obtain source catalogues in each of the five Hi-GAL wavebands. $\texttt{CuTEx}$ differentiates the image along 4 directions, $X$, $Y$ and the two diagonals twice. Negatives found in the second differential image are taken to be sources as these are the peaks against the background as long as they are coincident with zeroes in the first differential image. A source is identified when it is above the threshold in all 4 differential directions. The identified sources are then fitted with 2D elliptical Gaussian profiles, centred at the positions above to extract the integrated fluxes.

These infrared surveys and their extracted catalogues provide almost complete coverage of the mid-infrared to the sub-millimetre wavelength range, covering the whole SED of embedded YSOs.

\subsection{Bolocam Galactic Plane Survey}

The BGPS mapped the two regions, $\emph{l}$\,=\,30$\degr$ and $\emph{l}$\,=\,40$\degr$, as part of the 133\,deg$^{2}$ northern Galactic Plane section of the survey ($\emph{l}$\,= -10$\overset{\circ}{.}$5\,to\,90$\overset{\circ}{.}$5). The BGPS surveyed the continuum at 1.1 mm ($\nu$\,=\,271.1\,GHz), with an rms noise of 11\,--\,53\,mJy\,beam$^{-1}$ and an effective angular resolution of 33$\arcsec$.

\citet{Eden12,Eden13} derived distances, and as a result, masses and positions with respect to spiral arms or with Galactocentric radius to 989 catalogued sources \citep{Rosolowsky10}, 793 and 196 for the $\emph{l}$\,=\,30$\degr$ region and $\emph{l}$\,=\,40$\degr$ region respectively. The masses found for the BGPS sources were predominantly in the range of masses associated with molecular clumps (10$^{1.5}$\,--\,10$^{3}$\,M$_{\odot}$; \citealt{Dunham11}),  large, dense, bound regions within which stellar clusters and large systems form.

The 99 per cent completeness limits in the two fields are 450\,M$_{\odot}$ and 550\,M$_{\odot}$ at 14\,kpc in the $\emph{l}$\,=\,30$\degr$ and $\emph{l}$\,=\,40$\degr$ regions, respectively \citep{Rosolowsky10}.

\section{Source Luminosity Calculations}

\subsection{Infrared wavelength choice}

A YSO spectral energy distribution (SED) has two components (\citealt{Molinari08}; \citealt{Compiegne10}). The infrared part of the SED comes from the embedded YSO (if already nuclear burning) and/or the accretion luminosity, while the larger cold envelope within which the YSO/H\,{\sc ii} region is embedded emits strongly at the sub-millimetre and millimetre wavelengths. The latter component may be physically independent of the star-formation rate or star-forming content of a given clump until the clump material begins to be cleared in the later evolutionary stages \citep{Molinari08,Urquhart13b}.

By fitting an SED to the Hi-GAL data, the suitability of including the longer infrared and sub-millimetre wavelengths in a measurement of the YSO luminosity can be tested. These SEDs were fitted to the band-merged Hi-GAL sources that were positionally matched with the $\emph{l}$\,=\,40$\degr$ BGPS sources. The $\emph{l}$\,=\,40$\degr$ region was used as it has the lowest confusion levels of the two fields.

Before fitting a modified blackbody to the Hi-GAL SEDs, we scaled the SPIRE fluxes ($\ge$\,250\,$\upmu$m) by the ratio of the beam-deconvolved source size and beam size at the considered SPIRE wavelength (cf. \citealt{NguyenLuong11}). This corrects for the increase in the observed source size, and consequently the flux, at the longest $\emph{Herschel}$ wavelengths \citep[e.g.][]{Giannini12}, as well as for the larger beam size. The sub-millimetre fluxes are likely to be dominated by a cold envelope that is physically much larger than the embedded YSO. The longer-wavelength sources also tend to contain multiple shorter-wavelength detections within the larger beam.

The fitting strategy is that as described in \citet{Giannini12}, with the modified blackbody expressions and adapted constants the same as \citet{Elia13}. The free parameters are integrated flux, which is then converted to mass, $M$, and temperature, $T$, while the dust opacity exponent, $\beta$, is kept constant at 2. The fits do not include the 70\,$\upmu$m point as the greybody envelope may be contaminated by the emission of its protostellar content and this point generally lies above the fit.

\begin{figure}
\includegraphics[scale=0.50]{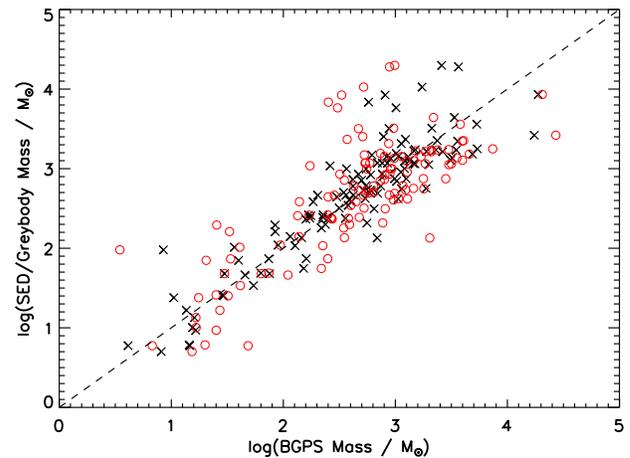}
\caption{Comparison of SED derived clump masses and masses calculated from the BGPS 1.1-mm flux density alone by \citet{Eden13}. The black crosses represent the BGPS masses calculated using the SED-derived temperature, whilst the red circles use a flat temperature of 14\,K. The 1:1 line is indicated.}
\label{fig:masscomparison}
\end{figure}

The masses derived from the SED fits were compared to those calculated from BGPS data, both using a fixed temperature of 14\,K (see figure 3 of \citealt{Eden13}) and with temperatures obtained from the SED fits. The result is shown in Fig.~\ref{fig:masscomparison}. The 1:1 ratio line is plotted over the distribution and it can be seen that the mass derived from the SED fitting is tracing the same mass as the 1.1-mm BGPS flux. This distribution also supports the use of a single temperature in the previous studies.

Since the two methods trace the same mass, it is safe to assume that the luminosity traced by the Hi-GAL data at wavelengths covered by the SED fit (160\,--\,500\,$\upmu$m) is dominated by the clump and the cold gas/envelope. This is not useful if trying to identify the clump-to-stars stage of the evolutionary sequence and measure its efficiency via $L/M$ as $M$ and $L$ are not independent, and if they are calculated from the same data, $M$ is essentially $L/T$. As a result, it is more pertinent to use shorter wavelength data, namely the 8 $\upmu$m GLIMPSE data, 12 and 22\,$\upmu$m WISE data and the 70\,$\upmu$m Hi-GAL data. These will trace the luminosity produced by the accretion disk and the embedded YSO, i.e. the stellar component, and the calculated luminosity and mass will be relatively independent. The wavelengths shorter than 8\,$\upmu$m are not used as they contribute very little to the luminosity.

\subsection{Source associations}

For a source to be considered a candidate YSO, it had to be catalogued in all 4 wavebands at 70\,$\upmu$m or shorter (i.e. 8, 12, 22 and 70\,$\upmu$m). The wavebands were combined by positional association. Starting at the largest beam size, an association in the waveband with the next largest was established if a source was found in that band within the beam radius. The order of the association was as follows: the WISE data at 6.5$^{\prime\prime}$ and 12$^{\prime\prime}$ for the 12 and 22\,$\upmu$m data respectively (the WISE all-sky survey contains an already band-merged catalogue, so only sources with a flux, and an error on the flux, were used), the 70\,$\upmu$m Hi-GAL data with an aperture of 4$^{\prime\prime}$ then the Spitzer GLIMPSE 8\,$\upmu$m with the smallest aperture of 1.9$^{\prime\prime}$. Table~\ref{sourcecounts} gives the numbers of sources at each wavelength. The matching process resulted in multiple associations as data with smaller beam sizes were introduced. It is usually assumed that only one of the smaller-beam-size sources is the ``true'' counterpart to the larger-beam-size components and the determination of the effect of an incorrect association where there is a choice is discussed in Section 3.4. A colour cut was applied that corresponded to the colour criteria used by the Red MSX Survey (RMS; \citealt{Lumsden13}), which was $F_{\rmn{22\,\mu m}}$ $>$ $F_{\rmn{8\,\mu m}}$. This limits the sample to the embedded YSO stage similar to RMS, resulting in an infrared selected sample of 1,641 sources, 1,050 and 591 in the $\emph{l}$\,=\,30$\degr$ and $\emph{l}$\,=\,40$\degr$ regions respectively.

\begin{table}
\begin{center}
\caption{Source counts at each wavelength in the $\emph{l}$\,=\,30$\degr$ and $\emph{l}$\,=\,40$\degr$ regions.}
\label{sourcecounts}
\begin{tabular}{lccc} \hline
Survey & Wavelength & $\emph{l}$\,=\,30$\degr$ & $\emph{l}$\,=\,40$\degr$ \\
& & Sources & Sources \\
\hline
GLIMPSE & 8\,$\upmu$m & 204,367 & 163,848 \\
WISE & 12 \& 22\,$\upmu$m & 57,506 & 47,179 \\
Hi-GAL & 70\,$\upmu$m & 4,578 & 1,469 \\
BGPS & 1.1\,mm & 806 & 229 \\ 
\hline
\end{tabular}
\end{center}
\end{table}

These associations were then positionally matched to the 1.1-mm BGPS sources, with only those sources that fell within the extent of a BGPS source counted, as catalogued by \citet{Rosolowsky10}. This allows the distance of the BGPS source to be assigned to the infrared (IR) source and only allow for IR sources to be counted which are part of a clump. These IR-BGPS associations resulted in a total of 241 infrared sources in the $\emph{l}$\,=\,30$\degr$ region and 107 in the $\emph{l}$\,=\,40$\degr$ region with assigned distances.

In total, 298 BGPS clumps are associated with YSOs, 29 per cent of the total 1,035 BGPS sources. \citet{Csengeri14} found a 30 per cent association when matching ATLASGAL and WISE sources. Of these 298 clumps, the vast majority are associated with a single infrared source (250 or 84 per cent), with only two clumps found to be associated with 3 infrared sources, one in each of the two regions.

To estimate the number of chance alignments, a Monte Carlo simulation was conducted, in which 1,641 sources were distributed over the $\emph{l}$\,=\,30$\degr$ and $\emph{l}$\,=\,40$\degr$ regions. The random distribution of simulated sources accounted for the true clustering scales of the IR sources by maintaining the relative position of sources found within a BGPS beam of each other. These sources were then matched with the BGPS source catalogue. 100 repeats found that the percentage of sources aligned with BGPS detections had a mean value of 0.8\,$\pm$\,0.3 per cent, compared to the actual association alignments of 14.7 per cent.

\subsection{Galactocentric radius distributions}

\begin{figure}
\begin{tabular}{l}
\includegraphics[scale=0.50]{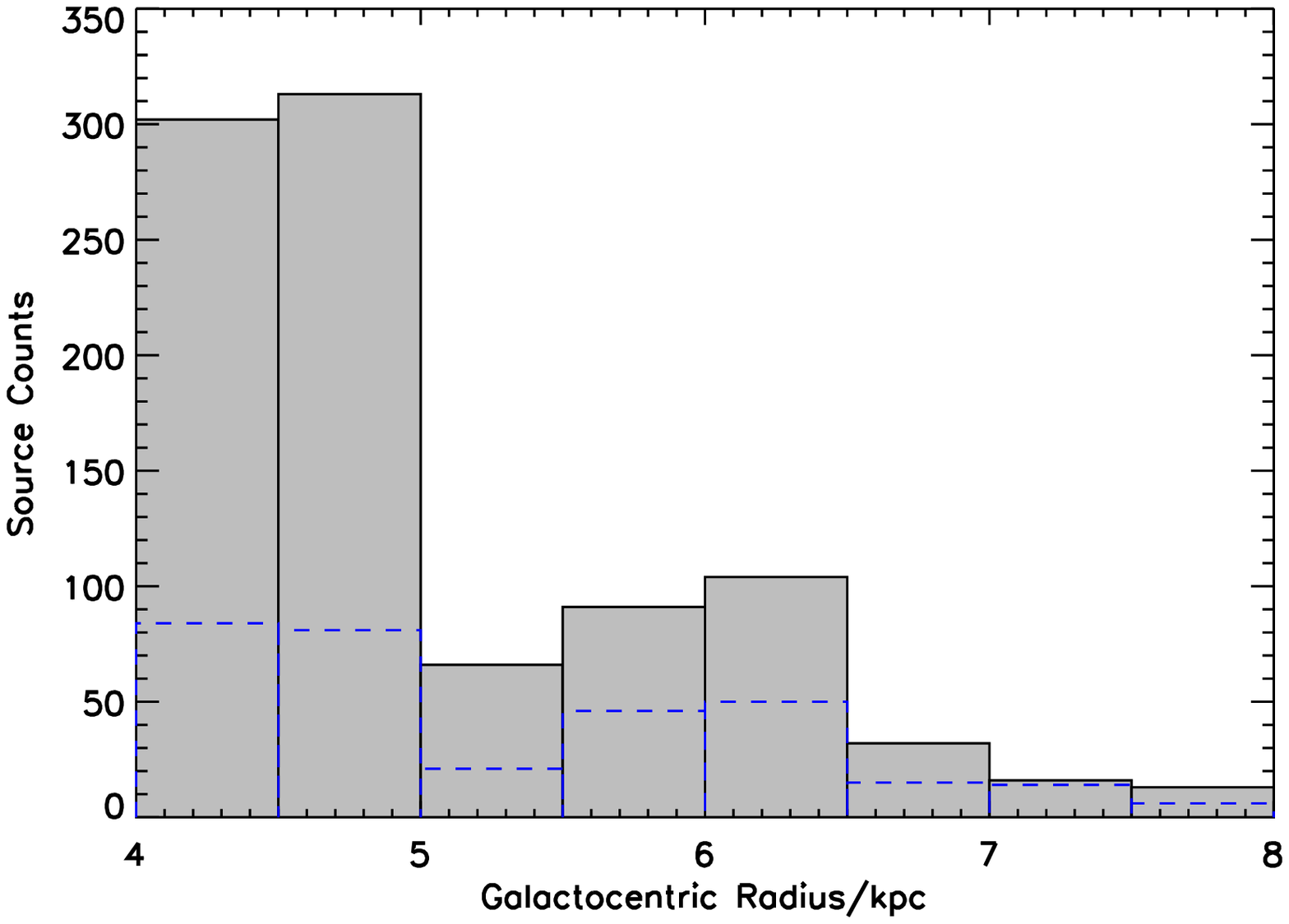}\\
\includegraphics[scale=0.50]{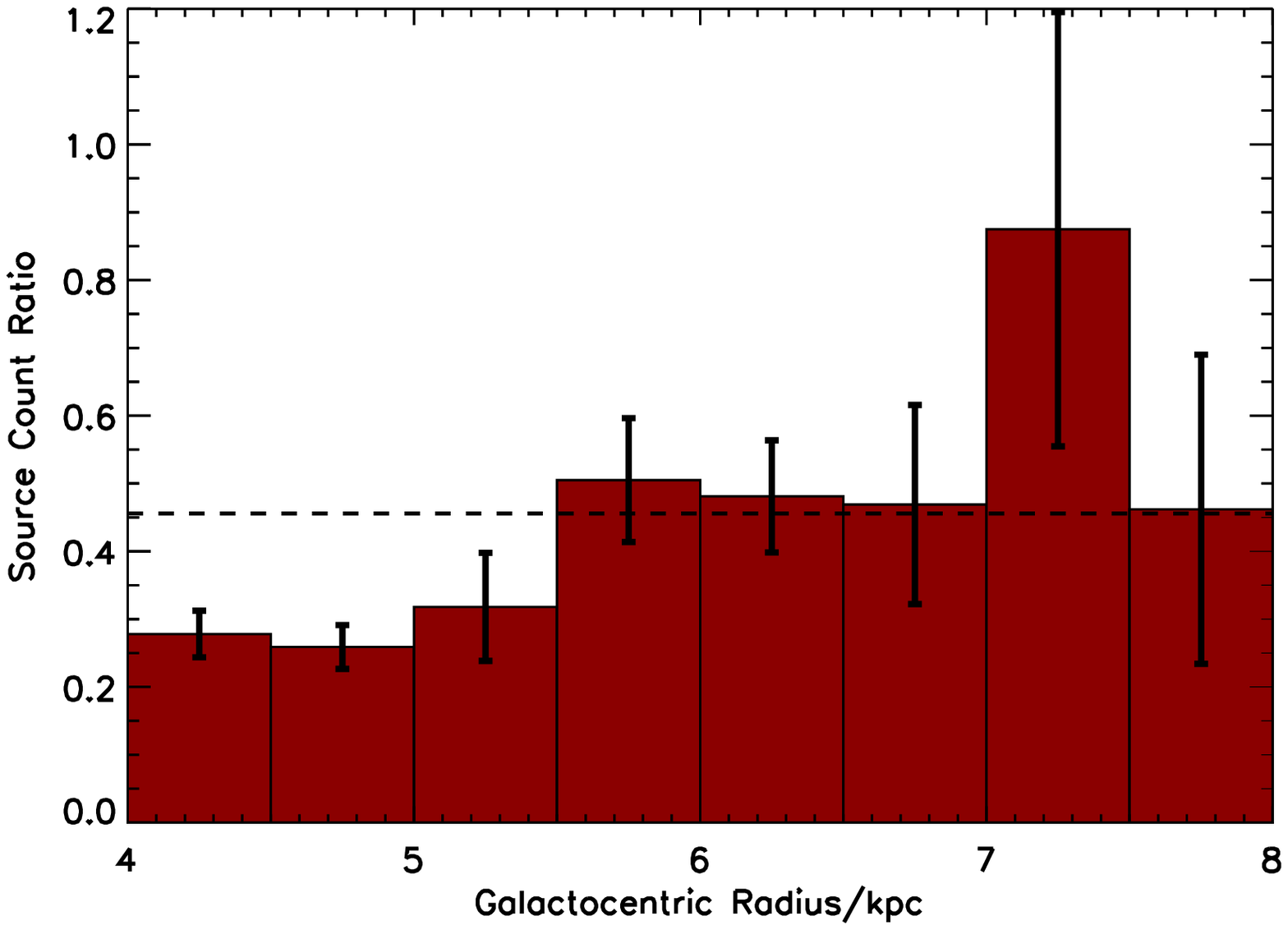}\\
\end{tabular}
\caption{Upper panel: Galactocentric radius distribution of the BGPS clumps (grey filled histogram) overlaid with the YSOs (blue dotted histogram). Lower panel: Ratio of the two distributions, with the mean value of 0.46 indicated by the dashed line. The errors are derived using Poissonian statistics.}
\label{GRSources}
\end{figure}

Using the assigned velocities and distances from \citet{Eden12,Eden13} for the BGPS clumps, and hence the YSOs, the distribution of the clumps and YSOs as a function of Galactocentric radius can be produced. These are displayed in the upper panel of Fig.~\ref{GRSources}. The ratio of these distributions is presented in the lower panel. This quantity gives the number of YSOs per dense clump and can be used as an analogue of the star-formation efficiency. The absolute luminosity-to-mass ratio as a function of Galactocentric radius is presented in Fig.~\ref{GRSFE} and discussed in Section 4.3. The peaks in both distributions occur at the positions of the Scutum--Centaurus tangent and W43. This is not surprising considering that the line of sight is down the spiral arm and W43 is the most prominent star-forming region in either of the lines of sight.

The number count ratio is relatively flat across most Galactocentric radii, with almost all the values for each 0.5\,kpc bin falling within 1- or 2-$\sigma$ of the mean. The mean value is found to be 0.46 with a standard deviation of 0.20 and a standard error on the mean of 0.07. The exceptions, however, are the bins at Galactocentric radii of 4-5\,kpc, with values $\sim$ 5-$\sigma$ below the mean. These radii are associated with the Scutum--Centaurus tangent as well as the W43 star-forming region. \citet*{Motte03} and \citet{NguyenLuong11} interpret this as evidence that the star-formation rate (SFR) will be higher in the future in W43 due to an excess of starless clumps.

\subsection{Luminosity calculations and the effect of incorrect association}

Two methods were used to calculate the YSO luminosities: an integration using the trapezium rule method and the online SED fitting tool of \citet{Robitaille07}. The latter makes use of a library of 200,000 models of YSOs. A detailed description of the models can be found in \citet{Robitaille06} and \citet{Robitaille07}. The trapezium integration assumes no physical model but may underestimate the total luminosity as there is no correction for the missing wavebands or contributions to the IR flux from the cold envelope component. The SED fits are based on models that may not be appropriate but provide the above extrapolation based on a spectral shape that is generally plausible.

\begin{figure}
\includegraphics[scale=0.50]{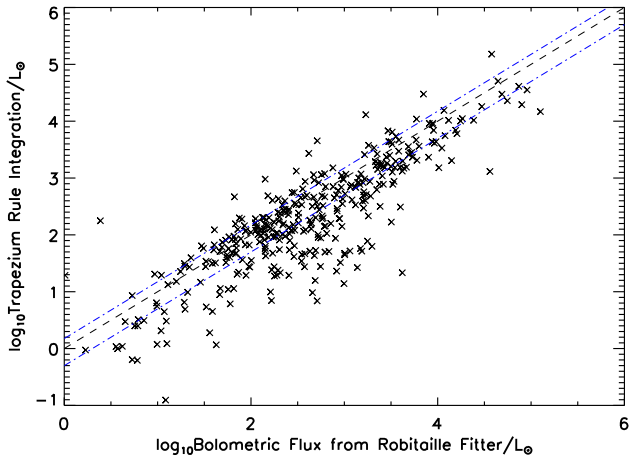}
\caption{The comparison of IR-BGPS sources showing the correlation between the trapezium rule integration and the Robitaille model luminosity. The 1:1 line is shown as the dashed line whilst the 50 per cent limits are the blue dot-dash lines.}
\label{RobTrapcorr}
\end{figure}

\begin{figure*}
\begin{tabular}{ll}
\includegraphics[width=0.48\textwidth]{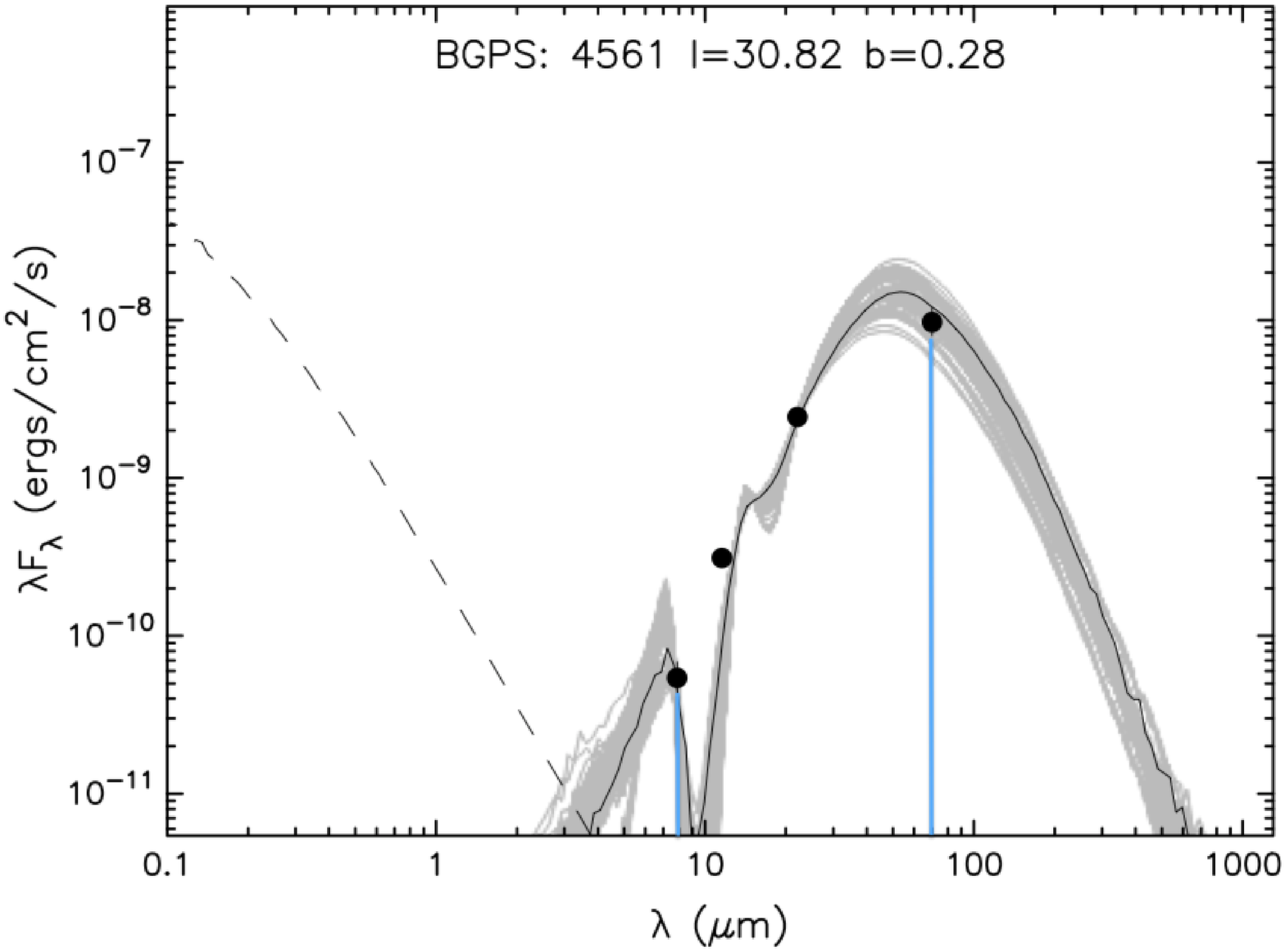} & \includegraphics[width=0.48\textwidth]{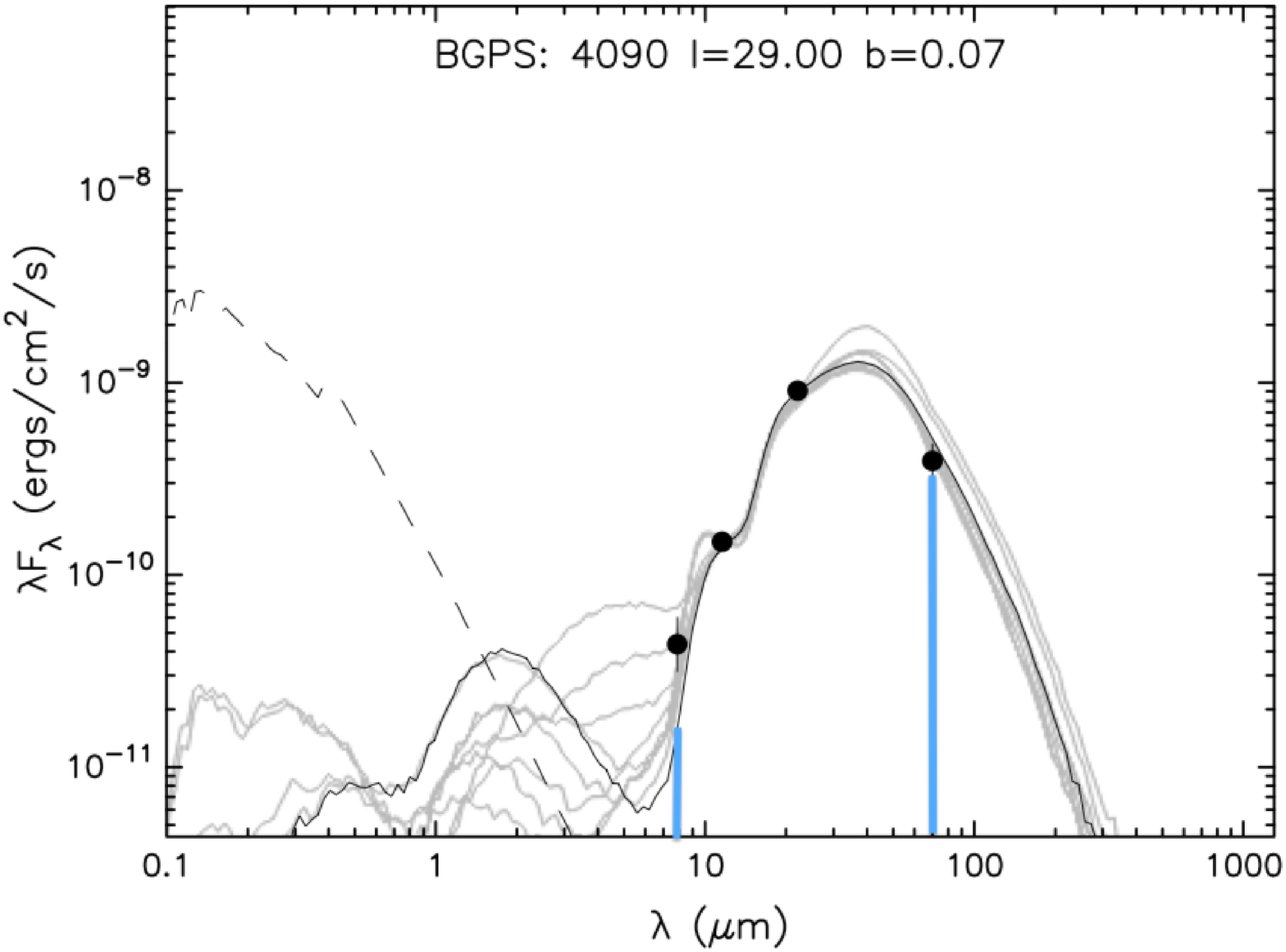} \\
\end{tabular}
\caption{Example Robitaille SEDs with the trapezium-rule integration limits marked by the two blue lines. The SED-fit luminosity is the total area under the curve. The black points are the data points to calculate the best fit, the black line represents the best fit with the next best 9 fits represented in grey. The dashed line shows the best model if a stellar object is assumed with no circumstellar extinction. Left panel: A source which exhibits similar luminosities for the Robitaille and trapezium-rule fits is shown with a trapezium-rule luminosity of 1.4\,$\times$\,10$^{4}$\,L$_{\odot}$ and Robitaille-fit luminosity of 2.4\,$\times$\,10$^{4}$\,L$_{\odot}$. Right panel: A source which exhibits an extreme difference in luminosity between the  Robitaille and trapezium-rule fits is shown with a trapezium-rule luminosity of 3.1\,$\times$\,10$^{2}$\,L$_{\odot}$ and Robitaille-fit luminosity of 4.0\,$\times$\,10$^{3}$\,L$_{\odot}$.}
\label{SEDs}
\end{figure*}

The distribution of the Robitaille SED calculated luminosities compared to the trapezium rule luminosities is shown in Fig.~\ref{RobTrapcorr}. The comparison shows the 50 per cent limits of the distribution of $L$ value ratios and 63 per cent of sources are found within these limits. The distribution is skewed to excess luminosity in the model SEDs. As can be seen from the right-hand panel of Fig.~\ref{SEDs}, SEDs fitted to these sources tend to have a significant component of emission at near-IR, or sub-millimetre, wavelengths, where we have no data constraints. This is symptomatic of the sources with a SED fit which overestimates the IR luminosity, a large component from either the sub-millimetre or near-IR. The sources with a greater trapezium integration are over-luminous at 70\,$\upmu$m. This makes the Robitaille fits unsuitable for those purposes as we are attempting to determine the IR luminosity independent of the sub-millimetre clump component. Fig.~\ref{SEDs} shows an example SED fit with a consistent trapezium-rule luminosity and one with a significantly overestimated Robitaille luminosity.

As a result, the luminosities of the YSOs used for this study were estimated via the trapezium rule. This however introduced a potential systematic error to the calculations since any emission at wavelengths $<$\,8\,$\upmu$m or $>$\,70\,$\upmu$m is not accounted for. However, the absolute values of the luminosity are less important than the comparative values and such biases are assumed to be similar across the sample. \citet{Veneziani13} estimated bolometric luminosities, rather than infrared luminosity, via a similar method to the trapezium rule, using those luminosities to produce star-formation rates and $L-M_{\rmn{clump}}$ relations for the two Hi-GAL science demonstration fields.

As mentioned above, multiple GLIMPSE and Hi-GAL sources were often found to be associated with a single WISE source, or vice versa, due to the different spatial resolutions. We therefore investigated the effect of making an incorrect choice of multiple components on the final calculated luminosity.

By fitting SEDs to all the potential infrared source matches for a given BGPS source in a subset of the sample, we investigated the uncertainty in luminosity values from the choice of source. Although some match combinations produced drastically better fits than others, the resulting luminosities were found to vary by less than 10 per cent. It is therefore clear that, for these purposes, there is no need to be overly concerned with identifying the correct counterpart and that the luminosity is dominated by wavebands (22\,$\upmu$m and 70\,$\upmu$m) in which there are few or no choices. Use of the trapezium rule produced similar results, with luminosities from each potential set of matches that were consistent within the measurement uncertainties. However, the absolute luminosities produced by the two methods were different.

\section{Results and Analysis}

\subsection{Star-formation efficiencies}

The star-formation efficiency, or SFE, considered here, one of several analogous variants, is the fraction of dense, dust-continuum traced clump material that has been converted into stars over some timescale. As with the clump-formation efficiency (CFE; the ratio of clump mass to cloud mass) analysis in \citet{Eden12, Eden13}, the SFE within the separate populations of clumps along the two lines of sight can be estimated. The SFE is the ratio of the luminosity produced by embedded young stellar objects to the mass of the clumps. 

The ratio of infrared luminosity of YSOs to clump mass, is a measure of:

\begin{equation}
\label{eq:SFE}
\frac{L_{\rmn{star}}}{M_{\rmn{clump}}}= \frac{1}{M_{\rmn{clump}}} \int^t_0 \frac{dL}{dt}\,dt
\end{equation}

\noindent where $dL/dt$ is the instantaneous star-formation rate in terms of $L$. A high value for $L/M$ can indicate either a high star-formation rate or a longer timescale. Equating $L/M$ with the SFE therefore requires a few assumptions. For precise linear dependence, $dL/dt$ must be proportional to $dM/dt$ which requires that the stellar IMF is fully sampled in all star-forming regions, up to the maximum stellar masses, i.e. ``sorted sampling'' filling the IMF \citep{Weidner06}. If, as seems likely, this is not the case and the IMF is filled stochastically \citep{Elmegreen06}, then $L/M$ still has a direct dependence on the SFE but it may be non-linear. Then, an increase in SFE must produce an increase in $L/M$ but an observed increase in $L/M$ can also be the result of the formation of a larger cluster at the same efficiency ($L$ $\propto$ $M^{2}$ for clusters). Such an increase in $L/M$ cannot be accounted for observationally until it is possible to count all the individual stars within a cluster. Another source of non-linearity may be introduced if the SFE is high, in which case, $M_{\rmn{clump}}$ is no longer constant, but SFE values appear to be generally less than 30 per cent (e.g. \citealt{Lada03}; \citealt{Johnston09}) and so $M_{\rmn{clump}}$ is taken to be constant over the lifetime of a star-forming event as traced by mid- and far-infrared detections. In principle, $L/M$ also increases with time and the SFE is usually defined with respect to some characteristic timescale; however, the stage of massive star formation traced by mid-infrared emission is short (less than a few $\times$ 10$^{5}$ years: \citealt{Davies11, Mottram11}) and our $L/M$ determination can be taken as a snapshot of the current SFE. Variations in $L/M$ can also be used as potential evolution indicators \citep{Molinari08, Urquhart14b} although $L$ and $M_{\rmn{clump}}$ tend to show a linear relationship with little evidence of evolution in $L/M$ during the IR-bright phase \citep{Urquhart14b}.

The clump populations can be divided into those belonging to the foreground spiral arms, the tangent region consisting of the Scutum--Centaurus tangent and the end of the long bar, and the background Sagittarius and Perseus arms in the $\emph{l}$\,=\,30$\degr$ region and the spiral arm and interarm regions in $\emph{l}$\,=\,40$\degr$. The $\emph{l}$\,=\,30$\degr$ region populations were separated using the distribution of the distances to the molecular clouds from \citet{Roman-Duval09}, as outlined in \citet{Eden12}. \citet{Eden13} details how the molecular clouds, and hence, the BGPS clumps were assigned to spiral arm and interarm populations in the $\emph{l}$\,=\,40$\degr$ region using the Galactic rotation model of \citet{Vallee95}.

Using the masses derived for the BGPS sources by \citet{Eden12, Eden13} and infrared luminosities for the associated YSOs obtained using the trapezium-rule integrations, we calculate the integrated luminosity and clump mass in each of the populations in the $\emph{l}$\,=\,30$\degr$ and $\emph{l}$\,=\,40$\degr$ fields and use the ratio, $L/M$ as an estimate of the SFE averaged over the scale of the field. The values for each region are displayed in Table~\ref{SFEs}. BGPS-traced clumps are included whether or not they contain IR sources since they contribute to the total $L/M$.

The results for the $\emph{l}$\,=\,30$\degr$ region differ from the results that are found in \citet{Eden12} who found consistent clump formation efficiencies in the tangent and background region and an increase in the foreground. The SFEs for each region depart from each other by factors of $\sim$ 2, that are significant within errors. The $\emph{l}$\,=\,40$\degr$ region shows that arm population has a value twice that of the interarm. The implications of this are discussed in Section 5.1.

\begin{table*}
\begin{center}
\caption{SFEs calculated from the YSO luminosity to clump mass ratio for the populations within the $\emph{l}$\,=\,30$\degr$ and $\emph{l}$\,=\,40$\degr$ regions.}
\label{SFEs}
\begin{tabular}{llccccc} \hline
Region & Population & $L_{\rmn{IR}}$ & No. of & $M_{\rmn{clump}}$ & No. of & SFE \\
& & (L$_{\odot}$) & YSOs & (M$_{\odot}$) & Clumps & (L$_{\odot}$/M$_{\odot}$) \\
\hline
& Foreground & 6750 $\pm$ 193 & 28 & 10300 $\pm$ 648 & 72 & 0.65 $\pm$ 0.05 \\
$\emph{l}$ = 30$^{\circ}$ & Tangent & 523000 $\pm$ 5090 & 160 & 544000 $\pm$ 9990 & 615 & 0.96 $\pm$ 0.02 \\
& Background & 492000 $\pm$ 6270 & 53 & 144000 $\pm$ 6880 & 106 & 2.47 $\pm$ 0.09 \\
\hline
$\emph{l}$ = 40$^{\circ}$ & Total Arm & 191000 $\pm$ 2570 & 77 & 137000 $\pm$ 9180 & 135 & 1.40 $\pm$ 0.10 \\
& Total Interarm & 59700 $\pm$ 2880 & 30 & 83300 $\pm$ 7070 & 61 & 0.72 $\pm$ 0.07 \\
\hline
\end{tabular}
\end{center}
\end{table*}

The uncertainties on the SFEs are a combination of the uncertainties in BGPS flux densities \citep{Rosolowsky10}, the variance of the temperature distribution of the source populations, the catalogued flux uncertainties on the GLIMPSE and WISE sources \citep{Churchwell09, Wright10}, as well as the recently released Hi-GAL catalogues (Molinari et al., in prep) containing the source flux uncertainties.

In summary, the average SFEs for the spiral arms are enhanced compared to the interarm regions by a factor of $\sim$ 2. However, there is a suppression towards the W43 star-forming region back to the interarm level. The other spiral arms, especially the Perseus arm, show an elevated SFE.

\subsection{Luminosity functions}

The luminosities of the YSOs were used to plot the luminosity functions (LFs) of the Scutum--Centaurus tangent region and of the other spiral arm regions. The spiral-arm subsample includes all YSOs found in the background region of $\emph{l}$\,=\,30$\degr$ and the Perseus and Sagittarius YSOs from $\emph{l}$\,=\,40$\degr$ as the background region consists of Sagittarius and Perseus YSOs. All sources with a heliocentric distance of less than 2\,kpc are omitted, in order to remove local sources that may contaminate the results and avoid including a large number of low-luminosity YSOs below the completeness limit at Galactocentric radii of 8\,kpc. No interarm LF is made due to the lack of sources.

\begin{figure}
\begin{tabular}{l}
\includegraphics[scale=0.50]{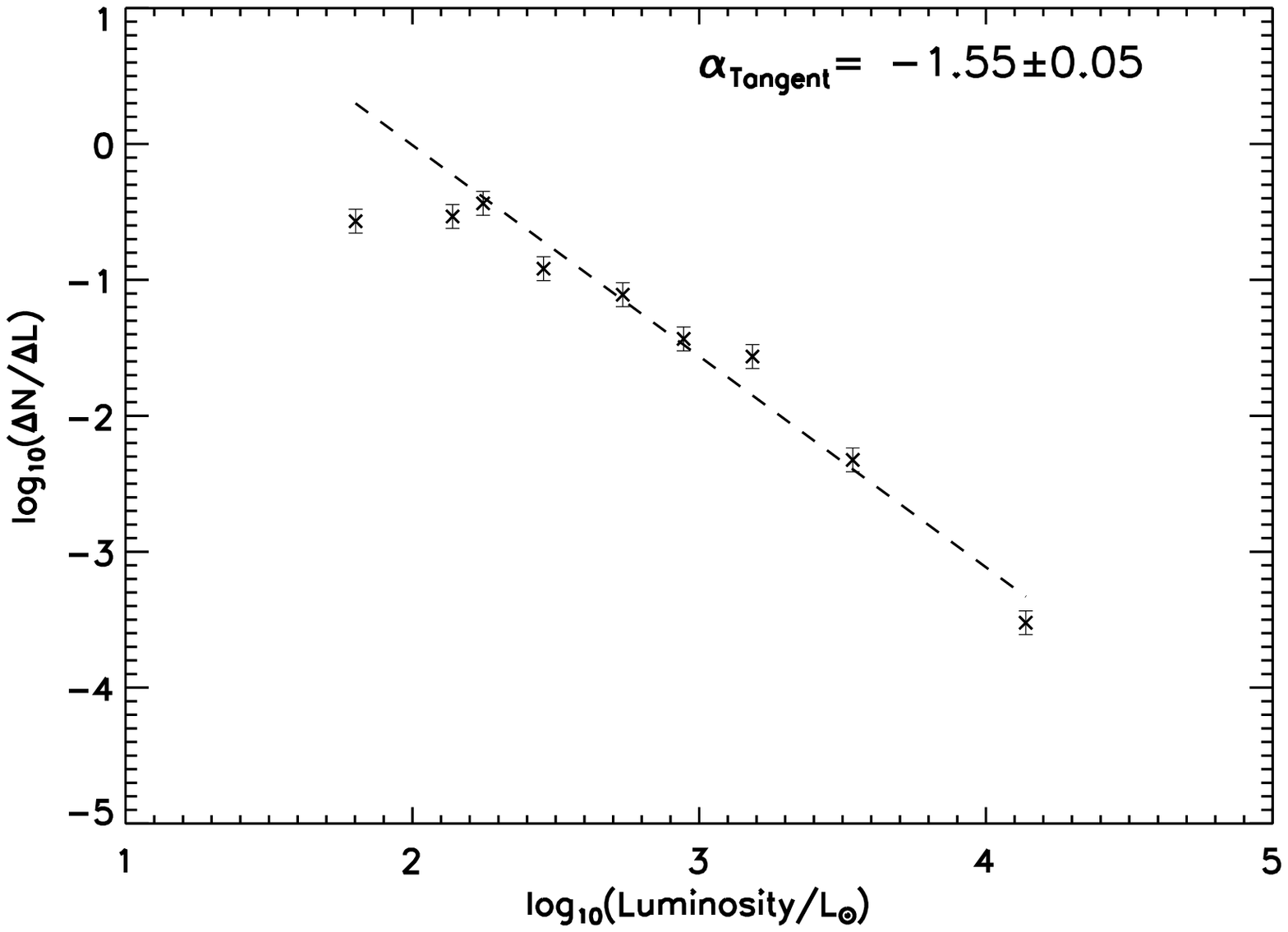}\\
\includegraphics[scale=0.50]{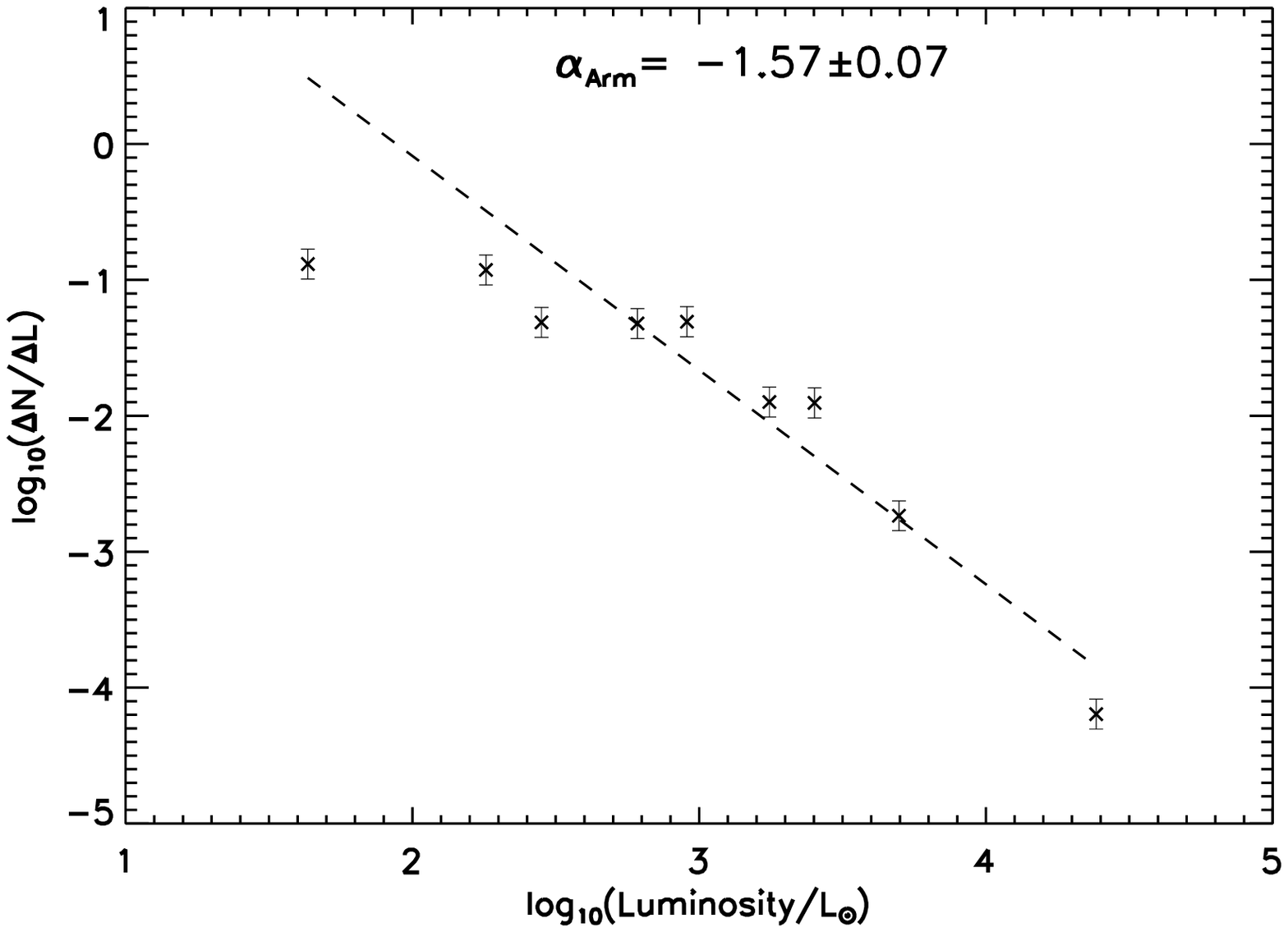}\\
\end{tabular}
\caption{LFs for the Scutum--Centaurus tangent region and all other spiral arm regions. The dashed lines are fits to the slope of the LF. The error bars correspond to $\sqrt{N}$ counting statistics. Top panel: Scutum--Centaurus region. Bottom panel: Spiral arm regions with sources of heliocentric distances less than 2\,kpc omitted from both the $\emph{l}$\,=\,30$\degr$ and $\emph{l}$\,=\,40$\degr$ regions.}
\label{LF}
\end{figure}

The LFs of these two populations are presented in Fig.~\ref{LF}. The plotted quantity, $\Delta\emph{N/}\Delta\emph{L}$, is the number of sources per unit luminosity interval in each luminosity bin, with the luminosity point representing the median value of the bin. A fixed number of sources per bin was used, as opposed to fixed bin widths, in order to equalise weights determined from Poisson errors.

If it is assumed that the data in both populations are distributed as a power law of the form $\Delta\emph{N/}\Delta\emph{L}$ $\propto$ $\emph{L}^{\alpha}$, a least-squares fit to each LF gives indices of $\alpha$\,=\,$-$1.55\,$\pm$\,0.05 and $\alpha$\,=\,$-$1.57\,$\pm$\,0.07 for the tangent and arm regions respectively. Both LFs begin to flatten at luminosities $\sim$ $log(L/$L$_{\odot})\,=\,2.5$ which we assume to indicate the sample completeness limit. The index values for these populations are consistent with each other, which shows that, in terms of the resulting distribution of YSO luminosities, the star-formation process at the end of the bar is no different to that in other spiral arms, despite the inclusion of the W43 region.

These slopes are consistent with that of the RMS survey \citep{Lumsden13}. Plotting the luminosities of both YSOs and H\,{\sc ii} regions \citep{Mottram11b,Urquhart14} with equal population bins give indices of $\alpha$\,=\,$-$1.48\,$\pm$\,0.04 and $\alpha$\,=\,$-$1.60\,$\pm$\,0.04 for the YSO and H\,{\sc ii} populations respectively. A combined sample had an index of $\alpha$\,=\,$-$1.50\,$\pm$\,0.02.

\begin{figure}
\includegraphics[scale=0.50]{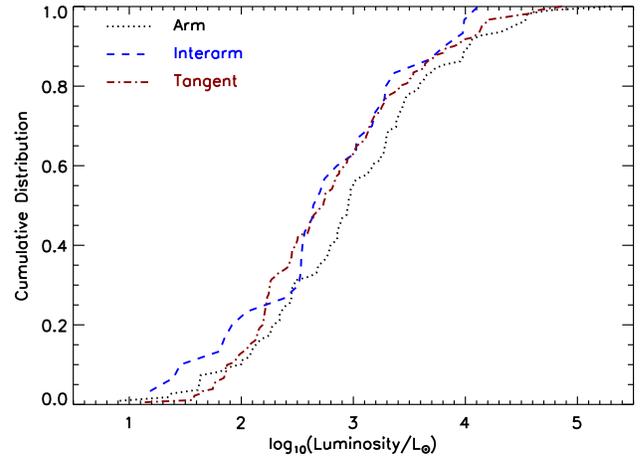}
\caption{The cumulative distribution function of the spiral arm, interarm and tangent region samples. The spiral arm sample is represented by the black dotted line, the interarm sample by the blue dashed line and the tangent by the red dot-dash line. The spiral arm sample again has sources of heliocentric distances less than 2\,kpc omitted.}
\label{cumulative}
\end{figure}

Further evidence for the similarity of the star formation in the two populations is that a Kolomogorov--Smirnov (K--S) test shows a 27 per cent probability that they are drawn from the same population, which is only a 1-$\sigma$ departure. Thus we are unable to reject the null hypothesis that they are drawn from the same population. Even though no LF was produced for the interarm regions, by performing a K--S test with the interarm sources, it showed a 23 per cent probability that they are drawn from the same population so there is no evidence for a variation in the current data. The cumulative distributions of the three samples are displayed in Fig.~\ref{cumulative}, and the similarity of the three distributions reflects the K--S test results. It is worth noting that the interarm sample is very small and not a strong constraint on the nature of interarm star formation.

\subsection{$L/M$ distribution with Galactocentric radius}

\begin{figure}
\includegraphics[scale=0.50]{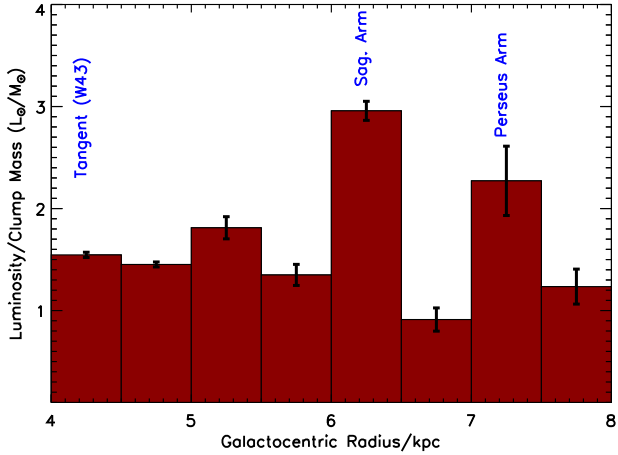}
\caption{The total $L/M$ as a function of Galactocentric radius within the studied regions centred at $\emph{l}$\,=\,30$\degr$ and $\emph{l}$\,=\,40$\degr$. The Galactocentric radii associated with spiral arms are labelled. Sources with heliocentric distances less than 2\,kpc are omitted, to remove local sources at Galactocentric radii of $\sim$ 8\,kpc.}
\label{GRSFE}
\end{figure}

The distribution of $L/M$ as a function of Galactocentric radius is displayed in Fig.~\ref{GRSFE}. All sources with heliocentric distances under 2\,kpc are again omitted. A significant peak is found at 6\,-\,6.5\,kpc, with an increase by factors of 2.2 and 3.2 over the adjacent bins, which correspond to 7.8-$\sigma$ and 9.3-$\sigma$ peaks, respectively. A less pronounced peak is found at 7\,-\,7.5\,kpc, with 4.1-$\sigma$ and 3.1-$\sigma$ departures found compared to adjacent bins, which are increases in factors of 2.5 and 1.9, respectively. The first peak corresponds to the Sagittarius spiral arm, whilst the second is the Perseus spiral arm.

\citet{Moore12} found peaks associated with the Sagittarius and Perseus arms in the $L/M_{\rmn{cloud}}$ distribution, using the CO-traced molecular cloud mass. These peaks were found to be extreme in their study due to the presence of the star-forming regions W51 and W49A, respectively. However, when they removed these star-forming regions, the Perseus arm had a SFE similar to the levels in the adjacent bins whereas the removal of W51 from the Sagittarius arm reduced the peak relative to the neighbouring bins but the peak remained.

This study has neither the W49 or W51 star-forming regions within the two lines of sight investigated. As a result, the distribution as seen in Fig.~\ref{GRSFE} should reflect that of \citet{Moore12}, once those regions are omitted as the $M_{\rmn{clump}}/M_{\rmn{cloud}}$ ratio is constant ($\sim$ 5 - 10 per cent) over large scales \citep{Eden12,Eden13}. The peak associated with the Sagittarius spiral arm is still elevated compared to the adjacent bins, whereas the Perseus arm bin at 7\,kpc has a marginal peak compared to the bins either side, hence reflecting the \citet{Moore12} distribution.

No peak is found at 4\,kpc, the radius associated with the Scutum--Centaurus tangent and the extreme-star-forming region, W43, suggesting that the current star formation in W43 is unspectacular in terms of the YSO luminosity being produced per unit gas mass.

\subsection{$L/M$ ratios of individual clumps}

The $L/M$ ratio of individual BGPS clumps can be calculated and the distribution of these values ranges from 0 to 140\,L$_{\odot}$/M$_{\odot}$. The mean value is 5.24\,$\pm$\,0.70\,L$_{\odot}$/M$_{\odot}$ with a median of 1.72\,$\pm$\,1.14\,L$_{\odot}$/M$_{\odot}$.

\begin{figure}
\includegraphics[scale=0.50]{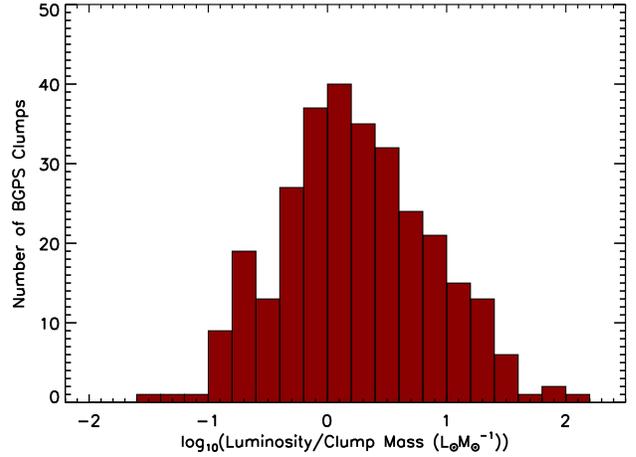}
\caption{The ratio of YSO luminosity to BGPS clump mass for all clumps with at least one associated YSO in the $\emph{l}$\,=\,30$\degr$ and $\emph{l}$\,=\,40$\degr$ regions. The bin size is 0.2\,dex.}
\label{normalSFE}
\end{figure}

The distributions of $L/M$ values in the separate regions listed in Table~\ref{SFEs} are statistically indistinguishable and can be assumed to be a single population. We can therefore combine these samples and the distribution of the total $L/M$ ratios for individual clumps is shown in Fig.~\ref{normalSFE}. The total distribution, as with the clump-formation efficiencies in \citet{Eden12, Eden13}, can be considered a lognormal distribution as Anderson--Darling and Shapiro--Wilk normality tests gave probabilities of 0.434 and 0.563, respectively, that the distribution is lognormal. The presence of a log-normal distribution indicates that the variation in $L/M$ is likely to be due to the random variations in a collection of individual processes that multiply together to produce the observed values, and this suggests that those values in the wings of the distribution, both with low and high $L/M$ are not the result of fundamentally different processes. No correlation of $L/M$ with clump mass was found. Therefore the spread of values in Fig.~\ref{normalSFE} is intrinsic and the result of natural variations in the star-formation process within clumps that are independent of this basic property.

The lognormal distribution can be fitted with a Gaussian that has a mean of 0.08 and a standard deviation of 0.57\,dex. With these parameters, three clumps can be found to have IR luminosity-clump mass ratios that are greater than the mean by more than 3-$\sigma$. With the probabilities of these events, only one source at these significances would be expected in a sample of this size. These three clumps are all found in the Scutum--Centaurus tangent in the $\emph{l}$\,=\,30$\degr$ region. These clumps have BGPS IDs of 4497, 4561 and 4617 with masses of 104, 197 and 97\,M$_{\odot}$, respectively. These all have one associated YSO which have luminosities of $log(L/$L$_{\odot})$\,=\,3.94, 4.15 and 4.13.

The absolute SFE values ($M_{\rmn{star}}/M_{\rmn{clump}}$) can be calculated for all the clumps which have an associated luminosity of $log(L/$L$_{\odot})$\,$>$\,3.01 by using the stellar luminosities outlined in \citet{Mottram11} for spectral types B3 and greater, calculated by using a 1\,Myr isochrone to local OB stellar populations using the stellar rotation models of \citet{Meynet03}. The range of these SFEs is from 0.1 to 14.3 per cent with a mean value of 1.9\,$\pm$\,0.3 per cent and a median of 0.9\,per cent, consistent with those of \citet{Smith14}. As these are most likely to be the sites of cluster formation, the SFEs calculated here are lower limits. This is only the star-formation efficiency derived from the brightest source and likely missing most of the lower-mass YSOs.

As we have determined the $L/M$ ratios of the individual clumps that contain a YSO, the range can be compared to those found for molecular clouds and whole spiral arms, to determine the scale at which the most drastic variations are occurring. The range of $L/M_{\rmn{clump}}$ for the individual clumps that have a YSO is 0.04\,$\pm$\,0.01\,--\,140\,$\pm$ 22\,L$_{\odot}$/M$_{\odot}$. The range of $L/M_{\rmn{cloud}}$ is 0.02\,$\pm$\,0.01\,--\,68\,$\pm$\,13\,L$_{\odot}$/M$_{\odot}$ and $L/M_{\rmn{arm}}$ is 1.00\,$\pm$\,0.02\,--\,2.9\,$\pm$\,0.2\,L$_{\odot}$/M$_{\odot}$.

\section{Discussion}

\subsection{Star formation on kiloparsec scales}

\citet{Eden12,Eden13} found that the dense-clump mass fraction in clouds ($M_{\rmn{clump}}/M_{\rmn{cloud}}$) varies by more than three orders of magnitude from cloud to cloud ($\sigma$ = 0.57) on spatial scales of 10-100\,pc. However, on kpc scales and between arm and interarm components, $M_{\rmn{clump}}/M_{\rmn{cloud}}$ is consistent with being constant, within the current uncertainties, and is seemingly unaffected by the presence of a spiral arm. \citet{Moore12} showed that $L/M_{\rmn{cloud}}$, averaged over kpc scales, increases a little in some spiral arms, relative to the nearby interarm regions, but not in others. In particular, the Scutum--Centaurus arm tangent shows no increase in $L/M_{\rmn{cloud}}$, despite being the location of a large enhancement in the surface density of the star-formation rate ($\Sigma_{\rmn{SFR}}$). Most of the large increases in $\Sigma_{\rmn{SFR}}$ associated with spiral arms are therefore due to crowding, rather than a change in the star-forming capacity of clouds. The residual enhancements in $L/M_{\rmn{cloud}}$ that were found in other arms are largely due to the inclusion of extreme star-forming regions such as W49A and W51 in the samples examined.

From the current work (Fig.~\ref{GRSFE}), we see that there are also increases in $L/M_{\rmn{clump}}$ associated at least with the Sagittarius arm, and possibly also with the Perseus arm, by factors of about 2, compared to the interarm regions, but, again, no enhancement in the Scutum--Centaurus tangent. Since we have evidence that $M_{\rmn{clump}}/M_{\rmn{cloud}}$ is constant on these scales, we expect that any variations in $L/M_{\rmn{clump}}$ in the present study would be consistent with those found in $L/M_{\rmn{cloud}}$ by \citet{Moore12} and, allowing for the differences in the samples used in the two studies (the current study does not include W49A, for example), this is indeed what we find.

\begin{figure}
\includegraphics[scale=0.50]{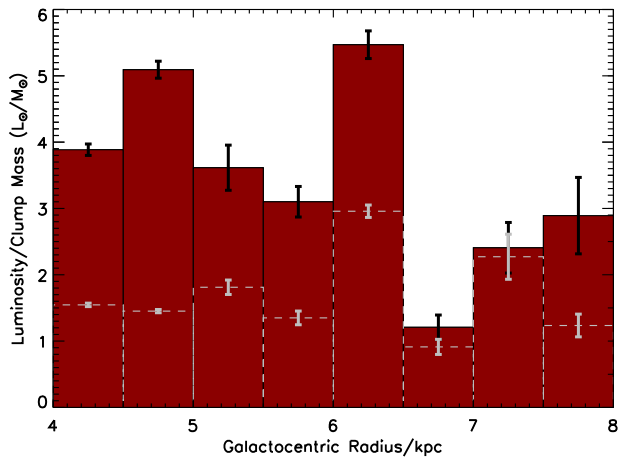}
\caption{The $L/M$ as a function of Galactocentric radius within the studied regions centred at $\emph{l}$\,=\,30$\degr$ and $\emph{l}$\,=\,40$\degr$ with the non-star-forming clumps removed. Sources with heliocentric distances less than 2\,kpc are omitted, to remove local sources at Galactocentric radii of $\sim$ 8\,kpc.}
\label{nostarless}
\end{figure}

The enhancements by a factor of $\sim$ 2 in the average $L/M_{\rmn{clump}}$ found in some spiral arms (Sagittarius and Perseus) but not in others (Scutum--Centaurus) might be explained by different time gradients in the star-formation rate in different places. If we remove the apparently non-star-forming clumps from the calculation of $L/M_{\rmn{clump}}$ (i.e. those clumps without an IR source), we find that a $\sim$ 30 per cent enhancement appears at the radius of the Scutum--Centaurus tangent (Fig.~\ref{nostarless}) with respect to the adjacent radii. $L/M_{\rmn{clump}}$ here is larger by a factor of $\sim$ 3 than the value obtained when the starless clumps are included in $M_{\rmn{clump}}$. Elsewhere within a radius of $\sim$ 6.5\,kpc, $L/M_{\rmn{clump}}$ increases by a factor closer to 2 than when starless clumps are not counted. This is consistent with the idea that the Scutum--Centaurus tangent region has a larger-than-average reservoir of dense clumps yet to form stars. \citet{NguyenLuong11} predict that the SFR in clouds associated with the W43 star-forming region, which is within the tangent and included in our sample, will be five times greater in the future than the current rate. Beyond 6.5\,kpc, and including the Perseus arm, the value of $L/M_{\rmn{clump}}$ is less dependent of the inclusion of starless clumps, indicating that there are relatively few of these.

In this scenario, the current gradient of the star-formation rate is relatively flat at radii less than 6.5 kpc, including the Sagittarius arm. In the Scutum--Centaurus tangent, the star-formation rate is set to increase with time and around the Perseus arm, star formation is currently relatively advanced and the production rate of clouds and of clumps within clouds is not keeping pace with the star formation. This picture is consistent with the known patchy nature of star formation in spiral arms and elsewhere \citep[e.g.][]{Urquhart14b} which naturally suggests that it is also intermittent in time. It may also require that the timescale for the infrared-traced star formation is longer than that for the production of dense clumps, so that clump production does not keep up, or that the dense clumps are aboriginal cloud features that are not replaced. It is usually assumed that since the timescales associated with the star-formation stages traced by infrared and sub-mm emission are a few $\times$\,10$^{5}$\,years or less (\citeauthor{Mottram11}, \citeyear{Mottram11}; \citeauthor{Ginsburg12}, \citeyear{Ginsburg12}), then these tracers give a snapshot of current star formation, with little or no look-back time.

As shown in \citet{Eden12,Eden13} and in Fig.~\ref{normalSFE}, the distributions of the clump-formation and star-formation efficiencies of different populations of individual clouds all follow statistically indistinguishable log-normal distributions. Since spiral arms generally have more star formation in them, the spiral-arm populations are larger and sample further into the wings of the $L/M_{\rmn{clump}}$ probability distribution, and so include more very high (and very low) SFE clouds and clumps but, since $L(M)$ for massive, star-forming clumps is linear rather than ZAMS-like \citep[e.g.][]{Urquhart14b}, the sample average of $L/M_{\rmn{clump}}$ does not depend on sample size and should not vary between arm and interarm populations for that reason. This does not rule out the possibility of some local variance produced by the inclusion of occasional extreme objects, as found in \citet{Moore12}.

A physical explanation as to why $L/M_{\rmn{clump}}$ should be generally enhanced in spiral-arm clumps is elusive. Star-formation timescales may be longer in longer-lasting and generally more massive clouds \citep{Dobbs11}, but there is no independent evidence to support this. Positive feedback (i.e. triggering) in crowded regions might be involved, although models suggest that this is not a significant effect \citep{Dale14} and there is no indication that clumps in spiral-arm clouds are physically different from those elsewhere. While a more definitive resolution of this question requires studies of larger samples, results so far contain no strong evidence of significant variations in SFE on kpc scales that cannot be ascribed to the inclusion of extreme objects in the samples and hence no indication that large-scale mechanisms in general, and spiral arms in particular, are important in regulating Galactic star formation. The fact that variations in conditions within individual clouds are much more significant, as outlined in Section 4.4, implicates cloud-scale mechanisms such as the initial conditions of cloud formation, cloud collisions or local feedback as the key factors.

\subsection{Potential sources of bias}

\subsubsection{Distance dependence of $L/M$ and completeness}

\begin{figure}
\includegraphics[scale=0.50]{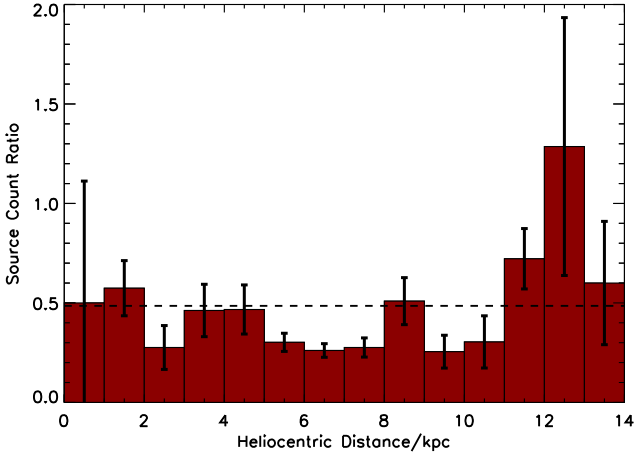}
\caption{Ratio of the number of YSOs to BGPS clumps as a function of heliocentric distance, with the mean value of 0.49 indicated by the dashed line. The errors are derived using Poissonian statistics.}
\label{DistSources}
\end{figure}

\begin{figure}
\includegraphics[scale=0.50]{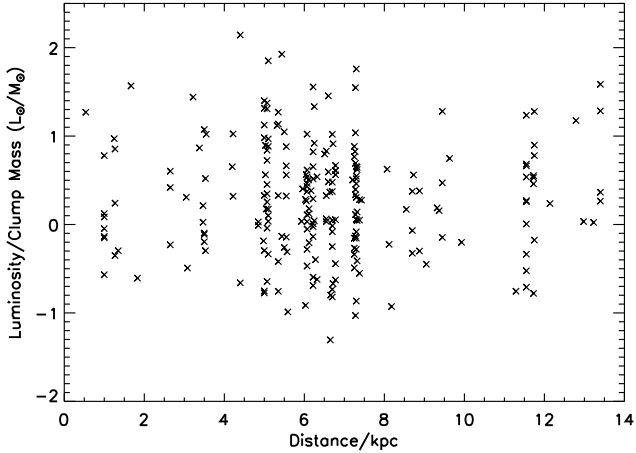}
\caption{$L/M$ ratio for individual clumps as a function of distance in the $\emph{l}$\,=\,30$\degr$ and $\emph{l}$\,=\,40$\degr$ regions.}
\label{complete}
\end{figure}

One source of potential bias relating to the $L/M$ ratio is the distance. There are three ways in which distance effects could alter the $L/M$ ratio: the nearby clumps have elevated $L/M$ ratios due to a greater detection percentage of YSOs, the most distant clouds having a deficit due to the YSO completeness dropping off before the BGPS completeness or the inverse of this and the lack of BGPS completeness pushing the $L/M$ ratio up.

The detection limits of the YSO-tracing surveys could simultaneously increase the $L/M$ ratios nearby and decrease it at further distances. A larger percentage of infrared sources would be observed in nearby clumps and clouds as the lower luminosity sources would have been detected. This effect would work in reverse at larger distances. This effect is not seen in Fig.~\ref{DistSources} which shows a flat distribution in the ratio of numbers of YSOs to clumps. The flatness of the distribution indicates that the surveys trace similar distances, with no enhancement or suppression of the ratio at the furthest distances. It is worth noting that any YSOs detected in nearby clumps, but missed at far distances, will be low luminosity and will not add much to $L/M_{\rmn{clump}}$. In the reverse case, undetected clumps at further distances will not be associated with any bright YSOs.

The respective completeness limits of the two surveys could also alter the result. The BGPS has a completeness limit of $\sim$ 550\,M$_{\odot}$ at 14\,kpc \citep{Rosolowsky10}, whilst the YSO luminosity is limited by the least sensitive survey, Hi-GAL 70\,$\upmu$m, but has a completeness limit at $\sim$ 675\,L$_{\odot}$ at 14\,kpc \citep{Elia13}. These surveys are similarly sensitive out to the furthest distances sampled in this study, and the lack of distance bias is shown in Fig.~\ref{complete}, with a flat distribution found.

\subsubsection{Area of Galaxy studied}

The regions sampled in this study are relatively narrow, in terms of the entire Galactic plane. However, it is unlikely that this introduces a significant bias as our results are fully consistent with that of \citet{Moore12} who found similar increases in $L/M_{\rmn{cloud}}$ over a much greater fraction of the Galaxy.

\subsubsection{Assumed spiral arm model}

For this work, and those of the clump-formation efficiency analysis \citep{Eden12,Eden13}, a four-armed model of the Galaxy is adopted. This model is preferred over contrasting two-arm models, determined from the kinematics of Two Micron All-Sky Survey (2MASS) sources \citep{Francis12}, or three-arm models from the distributions of giant molecular clouds \citep{Solomon87}.

The four-armed model is preferred due to the vast number of different tracers which outline four arms, from H\,{\sc ii} regions \citep[e.g][]{Paladini04}, to H\,{\sc i} distributions \citep{Levine06} to combinations of tracers \citep[e.g.][]{Hou09}. However, the YSO population of the Galaxy is consistent with a four-arm model \citep{Urquhart14}, and as we are tracking the YSO population, this is the best choice.

The choice of this model over others would have a bearing on the arm/interarm separation of the sources. It, however, would not have a bearing on the Galactocentric radius distribution. As such it would be clear that on the scale of individual clumps we would still see the large variation as seen in Fig.~\ref{normalSFE} but the variations on kiloparsec scales would be averaged out.

\subsection{Clump mass-luminosity relation}

\begin{figure}
\includegraphics[scale=0.50]{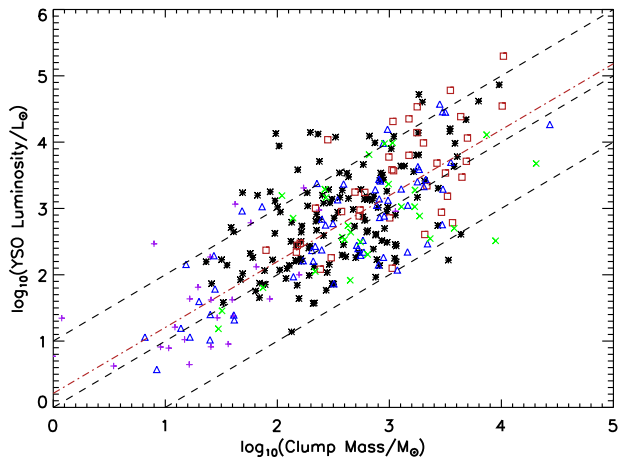}
\caption{Clump mass-luminosity relationship for the matched YSOs and BGPS clumps. The $\emph{l}$\,=\,30$\degr$ populations of the foreground, tangent and background regions are represented by purple plus symbols, black stars and red squares, respectively, with the $\emph{l}$\,=\,40$\degr$ arm and interarm sources represented by blue triangles and green crosses, respectively. The lower, middle and upper black dashed lines represent the $L/M_{\rmn{clump}}$ = 0.1, 1 and 10\,$L_{\odot}/M_{\odot}$, respectively. The red dot-dash line represents the fit to the data.}
\label{M/L}
\end{figure}

The relationship between clump mass and YSO luminosity is displayed in Fig.~\ref{M/L}. There appears to be a correlation, but both quantities have a strong distance dependence. Therefore, to test the correlation, a partial Spearman correlation test, which takes into account the dependence on a common variable (distance in this case), was used. This gave a correlation coefficient of 0.56 with a $p$-value of $\sim$ $\times$ 10$^{-7}$ that this correlation occurred by chance for the sample of 298 BGPS clumps. By selecting a distance-limited sample between 4 and 6\,kpc, a correlation coefficient of 0.52 with a $p$-value of $\sim$ $\times$ 10$^{-5}$ is found, therefore there is no evidence for any significant systematic distance effect on the observed relationship.

Interpretation of the $L-M_{\rmn{clump}}$ relation is difficult if the $L$ and $M_{\rmn{clump}}$ are derived from the same data \citep[e.g.][]{Veneziani13}, which is particularly the case if $L$ = $L_{\rmn{bol}}$, hence the use of $L_{\rmn{IR}}$ instead of $L_{\rmn{bol}}$ to attempt to remove the dependence of both quantities on the sub-mm component of the SED. The linear log-log fit of $L_{\rmn{IR}}$ with $L_{\rmn{bol}}$ in Fig.~\ref{RobTrapcorr} is the result of YSO SEDs being essentially greybody spectra and that the sources are selected in the same way i.e. by waveband, with small variations, such as the temperature differences reflected in the differing colours of the sources and the independence of $L_{\rmn{IR}}$ and $L_{\rmn{bol}}$ causing the scatter.

The slope of the linear log-log fit found for this distribution is 0.99\,$\pm$\,0.04, which is slightly shallower than that of \citet{Molinari08}, but is consistent with that of \citet{Urquhart14b} who found slopes of 1.27\,$\pm$\,0.08 and 1.09\,$\pm$\,0.13, respectively. The slope of the linear fit to the distance-selected sample mentioned above is 0.89\,$\pm$\,0.18.

Further to this correlation, the $L-M_{clump}$ relation can be used to derive the evolutionary state of the YSO as it approaches the main sequence (\citealt{Molinari08}; \citealt{Giannetti13}). \citet{Molinari08} highlight two phases in this evolution in the $L-M_{\rmn{clump}}$ plane: the main accretion and envelope clean-up phases, using the accelerating accretion rates as postulated by \citep{McKee03}. Once the clump is formed, the gradient of the locus in $L-M_{\rmn{clump}}$ space will begin at $L_{\rmn{IR}}$ = 0, as there would be no YSOs present, and so $L_{\rmn{IR}}$ will be zero. The gradient will be 1 if $L$\,=\,$L_{\rmn{bol}}$ and the temperature is constant with $L$, and $M$. As YSOs form, the luminosity will be made up of two components, the accretion luminosity and the luminosity of the source. If the accretion luminosity, $L_{\rmn{acc}}$, dominates the YSO luminosity, then this implies a high accretion rate and/or a very early stage prior to fusion. The luminosity thus increases at near-constant mass on a vertical track in $L-M_{\rmn{clump}}$ space. Once the accretion finishes, the envelope begins to be dispersed. The YSOs find themselves clustering at $L/M_{\rmn{clump}}$\,=\,1\,L$_{\odot}$/M$_{\odot}$ as sources spend a relatively long time in the region where accretion is ending and envelope clearance has not yet commenced. Both the accretion rate and envelope clean-up rate increase with time with the accretion rate increasing as the mass of the protostar increases (\citealt{McKee03}; \citealt{Davies11}). The clean-up phase starts slowly as it takes some time for the embedded star to disrupt the clump from which it formed. The final luminosity is dependent on the initial clump mass, and as a result a linear log-log distribution is found, especially as the accretion luminosity is proportional to the stellar mass \citep{McKee03}. This dependence, $L$\,$\propto$\,$M^{0.95}_{*}$, implies the accretion luminosity is proportional to the initial clump mass due to the relationship between stellar mass and clump mass \citep{McKee03}.

Once the clearing is finished, the YSOs move to the left in $L-M_{\rmn{clump}}$ space, which is a very quick phase, hence why no very-high-luminosity, low-mass sources are found. By selecting sources in the mid-IR, we may be filtering out the envelope-clearing and pre-stellar sources. This begins the evolution from the slope of unity towards the mass-luminosity relation of either main-sequence stars ($L$\,$\propto$\,$M^{3.5}$) or clusters ($L$ $\propto$ $M^{2}$).

The distribution also implies that there is a constant $L/M_{\rmn{clump}}$ ratio acting along the spectrum of clump masses. It however, requires that the instantaneous SFE, i.e. the mass of stars formed from a particular mass of dense gas, acting on each clump varies as a function of clump mass \citep{Urquhart14b}. \citep{Urquhart14b} suggest that the SFE is lower for the high-mass and high-luminosity sources in which the high-mass members of a cluster form first. The disparity between the observed and cluster relationships is seen by both \citet{Sridharan02} and \citet{Urquhart14b}. Both of those studies show that the lower mass clumps are over luminous, whilst the higher mass clumps are not as luminous as one would expect if each source was expected to be the site of cluster formation and a standard IMF was sampled at a constant SFE. However, the observed dependence is consistent with the relationship between accretion luminosity and stellar mass ($L$\,$\propto$\,$M^{0.95}$) and that of accretion luminosity and final stellar mass ($L$\,$\propto$\,$M^{1.20}$; \citealt{McKee03}). This implies that these sources are either still accreting or have just reached the final stellar mass.

\section{Summary \& Conclusions}

We have combined a series of surveys tracing infrared YSOs with the Bolocam Galactic Plane Survey (BGPS) sample of star-forming clumps within two slices of the Galactic plane, in order to investigate how the star-formation efficiency, measured as the ratio of infrared luminosity to clump mass, varies with Galactic environment. The two slices of the Galactic plane, centred at Galactic longitudes of $\emph{l}$\,=\,30$\degr$ and $\emph{l}$\,=\,40$\degr$, cover spiral-arm regions, interarm regions and the end of the Galactic Long Bar, which contains the star-forming region, W43.

As a result of positionally matching infrared sources to BGPS clumps using the sample subsets from \citet{Eden12, Eden13}, we find a total of 348 candidate YSOs associated with 298 distinct clumps, which accounts for 29 per cent of the BGPS sources found in the two regions. The distances, and Galactic environments, of the BGPS sources and the luminosities of the associated YSO sources were taken from the derived distances in \citet{Eden12, Eden13}.

In order to obtain a YSO luminosity independent of the clump mass, we used only wavelengths shorter than 70\,$\upmu$m to calculate $L_{\rmn{IR}}$ and the 1.1-mm flux density for $M_{\rmn{clump}}$. The method for calculating the luminosities was to use a trapezium rules integration in that wavelength range as the use of the bolometric luminosity from the Robitaille SED fitting tools \citep{Robitaille06, Robitaille07} included the contribution from the envelope in which the YSO is embedded, i.e. the clump, and an independent measure of the YSO luminosity from the clump component was required. The absolute luminosities found using both methods were found to be different but the relative differences were consistent across all distances and environments.

We compared the ratio of infrared luminosity, $L_{\rmn{IR}}$, to the clump mass, $M_{\rmn{clump}}$, as a function of Galactic environment, and the properties of clumps hosting a candidate YSO and our main findings are:

\begin{enumerate}[(i)]

\item In the $\emph{l}$\,=\,30$\degr$ region, the $L_{\rmn{IR}}/M_{\rmn{clump}}$ ratio is larger in the background spiral arms by a factor of $\sim$ 2.5 compared to the other $\emph{l}$\,=\,30$\degr$ populations, including the W43 star-forming region. This is consistent with the results of \citet{Moore12} who found that the Sagittarius and Perseus arms had a higher $L_{\rmn{IR}}/M_{\rmn{cloud}}$ ratio than the Scutum--Centaurus arm. This may imply that the future SFR of W43 will be higher than the current observed rate, as suggested by \citet{NguyenLuong11}.

\item The luminosity distributions for the Scutum--Centaurus tangent region and a combined spiral-arm sample, made up of Sagittarius and Perseus arm sources, are consistent with each other. This supports the $M_{\rmn{clump}}/M_{\rmn{cloud}}$ results of \citet{Eden12} that the mass function at the end of the bar is not significantly different from any other spiral arm in the Galaxy and that there is no enhancement in the SFR per clump mass due to the interaction of the bar with spiral arms.

\item The observed distribution of $L_{\rmn{IR}}/M_{\rmn{clump}}$ ratios for all clumps is consistent with a log-normal probability function, implying that those with extreme values, both low and high, are the result of the same combination of random processes and do not indicate the presence of fundamentally different conditions.

\item Assuming a single star forming from each YSO, the average absolute SFE, ratio of stellar mass to clump mass, was found to be 1.9\,$\pm$\,0.2 per cent.

\item The ranges of $L/M$ found for the individual clumps, molecular clouds and spiral arms were found to be 0.04\,$\pm$\,0.01\,--\,140\,$\pm$\,22, 0.02\,$\pm$\,0.01\,--\,68\,$\pm$\,13 and 1.00\,$\pm$\,0.02\,--\,2.9\,$\pm$\,0.2\,L$_{\odot}$/M$_{\odot}$ respectively.

\item Combining these results with others investigating the role of spiral arms on the star-formation process it is found that, on kiloparsec scales, there is little evidence of significant variations in SFE that can be attributed to the action of the arms on the ISM. Differences in average $L_{\rmn{IR}}/M_{\rmn{clump}}$ found in spiral-arm sources may be due to gradients in the SFR or local variance in the limited samples. The most significant variations in SFE are found to be on the scales of individual molecular clouds and clumps.

\item There is a linear relationship between the  $L_{\rmn{IR}}$ and $M_{\rmn{clump}}$ with a best-fit power-law exponent of 0.99\,$\pm$\,0.04, which is consistent with that of \citet{Molinari08} and \citet{Urquhart14b}.

\end{enumerate}

\section*{Acknowledgments}

The authors wish to thank the anonymous referee for comments that have improved the quality of the paper. This work is based in part on observations made with the $\emph{Spitzer Space Telescope}$, which is operated by the Jet Propulsion Laboratory, California Institute of Technology under a contract with NASA. This publication makes use of data products from the Wide-field Infrared Survey Explorer, which is a joint project of the University of California, Los Angeles, and the Jet Propulsion Laboratory/California Institute of Technology, funded by the National Aeronautics and Space Administration. DJE wishes to acknowledge an STFC PhD studentship for this work. This research has made use of NASA's Astrophysics Data System.

\bibliographystyle{mn2e}
\bibliography{ref}

\label{lastpage}

\end{document}